\documentclass[prl,twocolumn, superscriptaddress,10pt]{revtex4-1}

\usepackage{packages}
\usepackage{notations}

\tikzset{
    partial ellipse/.style args={#1:#2:#3}{
        insert path={+ (#1:#3) arc (#1:#2:#3)}
    }
}

\tikzset{middlearrow/.style={
        decoration={markings,
            mark= at position 0.55 with {\arrow[thick]{#1}} ,
        },
        postaction={decorate}
    }
}

\def\empt{\tikz[baseline=-0.5ex,every node/.style={scale=0.9, inner sep=0pt}]{
\node[state,minimum size=5pt] (0) {~};
}}

\def\full{\tikz[baseline=-0.5ex,every node/.style={scale=0.9, inner sep=0pt}]{
\node[state,fill=black,minimum size=5pt] (0) {~};
}}

\def\emptempt{\tikz[baseline=-0.5ex,every node/.style={scale=1, inner sep=0pt}]{
\node[state,minimum size=5pt] (0) {~};
\node[state,right=6pt of 0,minimum size=5pt]                    (1) {~};}
}

\def\emptfull{\tikz[baseline=-0.5ex,every node/.style={scale=1, inner sep=0pt}]{
\node[state,minimum size=5pt] (0) {~};
\node[state,fill=black,right=6pt of 0,minimum size=5pt]                    (1) {~};}
}

\def\fullempt{\tikz[baseline=-0.5ex,every node/.style={scale=1, inner sep=0pt}]{
\node[state,fill=black,minimum size=5pt] (0) {~};
\node[state,right=6pt of 0,minimum size=5pt]                    (1) {~};}
}

\def\fullfull{\tikz[baseline=-0.5ex,every node/.style={scale=1, inner sep=0pt}]{
\node[state,fill=black,minimum size=5pt] (0) {~};
\node[state,fill=black,right=6pt of 0,minimum size=5pt]                    (1) {~};}
}

\newcommand{\lattice}{
\begin{tikzpicture}[scale=1, every node/.style={scale=1, inner sep=0pt}]

\node[state,fill=black,minimum size=5pt]                                   (0) {~};
\node[state,right=10pt of 0,minimum size=5pt]                    (1) {~};
\node[state,fill=black,right=10pt of 1,minimum size=5pt]                    (2) {~};
\node[state,fill=black, right=10pt of 2,minimum size=5pt]                    (3) {~};
\node[state,right=10pt of 3,minimum size=5pt]                    (5) {~};
\node[state,fill=black,right=10pt of 5,minimum size=5pt]                    (6) {~};
\node[state, right=10pt of 6,minimum size=5pt]                    (7) {~};
\node[state, right=10pt of 7,minimum size=5pt]                    (8) {~};
\node[state, fill=black, right=10pt of 8,minimum size=5pt]                    (9) {~};

\node[left=20pt of 0]                                   (left-2) {\dots};
\node[right=20pt of 9]                                   (right-1) {\ldots};

\node[above= 30pt of left-2] (deli-1){~};
\node[above= 30pt  of right-1] (deli-2){~};

\draw[dotted,-latex] (deli-1) --  (deli-2) ;

\pgfmathsetmacro{\slope}{0.5pt}

\pgfmathsetmacro{\basisHeight}{50}

  \foreach \x in {0,1,2,3,5,6,7,8,9}
    {        
           \draw ($ (\x)+(0,28pt)$) -- ($ (\x)+(0,34pt)$);
    }

 \draw[semithick] ($ (0)+(-15pt,\basisHeight pt)$) -- ($ (0)+(0,\basisHeight pt)+(0,\slope)$);
  \draw[semithick]  ($ (0)+(0,\basisHeight pt)+(0,\slope)$) -- ($ (1)+(0,\basisHeight pt)-(0,0*\slope)$) ;
    \draw[semithick]   ($ (1)+(0,\basisHeight pt)+(0,0*\slope)$) --  ($ (2)+(0,\basisHeight pt)+(0,\slope)$) ;
    \draw[semithick]   ($ (3)+(0,\basisHeight pt)+(0,2*\slope)$) --  ($ (2)+(0,\basisHeight pt)+(0,\slope)$) ;
        \draw[semithick]   ($ (3)+(0,\basisHeight pt)+(0,2*\slope)$) --  ($ (5)+(0,\basisHeight pt)+(0,1*\slope)$) ;
                \draw[semithick]   ($ (6)+(0,\basisHeight pt)+(0,2*\slope)$) --  ($ (5)+(0,\basisHeight pt)+(0,1*\slope)$) ;
   \draw[semithick]   ($ (6)+(0,\basisHeight pt)+(0,2*\slope)$) --  ($ (7)+(0,\basisHeight pt)+(0,1*\slope)$) ;
     \draw[semithick]   ($ (8)+(0,\basisHeight pt)+(0,0*\slope)$) --  ($ (7)+(0,\basisHeight pt)+(0,1*\slope)$) ;
          \draw[semithick]   ($ (8)+(0,\basisHeight pt)+(0,0*\slope)$) --  ($ (9)+(0,\basisHeight pt)+(0,1*\slope)$) ;

 \draw[dashed,semithick]   ($ (6)+(0,\basisHeight pt)+(0,0*\slope)$) --  ($ (5)+(0,\basisHeight pt)+(0,1*\slope)$) ;
   \draw[dashed,semithick]   ($ (6)+(0,\basisHeight pt)+(0,0*\slope)$) --  ($ (7)+(0,\basisHeight pt)+(0,1*\slope)$) ;
   
   \draw[latex -,shorten >=6pt,shorten <=6pt] ($ (6)+(0,\basisHeight pt)+(0,0*\slope)$) --($ (6)+(0,\basisHeight pt)+(0,2*\slope)$) ;
   \node[above=1.6*\basisHeight  pt of 6]      (60) {$\rmd W^j_t$};
            \node[below=5  pt of 6]      (600) {$j'$};   
   
   \draw[dashed,semithick]  ($ (0)+(0,\basisHeight pt)+(0,\slope)$) -- ($ (1)+(0,\basisHeight pt)+(0,2*\slope)$) ;
    \draw[dashed,semithick]   ($ (1)+(0,\basisHeight pt)+(0,2*\slope)$) --  ($ (2)+(0,\basisHeight pt)+(0,\slope)$) ;  

   \draw[ -latex,shorten >=6pt,shorten <=6pt] ($ (1)+(0,\basisHeight pt)+(0,0*\slope)$) --($ (1)+(0,\basisHeight pt)+(0,2*\slope)$) ;
      \node[above=1.6*\basisHeight  pt of 1]      (10) {$\rmd \overline{W}^{j'}_t$};
            \node[below=5  pt of 1]      (100) {$j'$};

\path (1.90) edge[thick,bend right=-50,latex'-] (2.90);
\path (6.90) edge[thick,bend right=-50,-latex'] (7.90);

\end{tikzpicture}
}

\newcommand{\latticeQuantum}{
\begin{tikzpicture}[scale=1, every node/.style={scale=1}]
\node[state,fill=black,minimum size=4pt]                    (0) {~};
\node[above=3pt] at (0.north) {$ j-1 $};
\node[state,fill=black,right=50pt of 0,minimum size=4pt]                    (1) {~};
\node[above=3pt] at (1.north) {$ j $};
\node[state,fill=black, right=50pt of 1,minimum size=4pt]                    (2) {~};
\node[above=3pt] at (2.north) {$ j+1 $};
\node[state,fill=black, right=50pt of 2,minimum size=4pt]                    (3) {~};
\node[above=3pt] at (3.north) {$ j+2 $};

\path (1.300) edge[bend right=40,-latex] (2.240);
\node[below=9pt] at ($0.5*(1.south)+0.5*(2.south)$.north) {$ \rmd W_t^j $};

\path (1.60) edge[bend right=-40,latex-] (2.120);
\node[above=9pt] at ($0.5*(1.north)+0.5*(2.north)$.north) {$ \rmd \overline{W}_t^j $};

\path (0.300) edge[dotted,bend right=40,-latex] (1.240);
\path (0.60) edge[dotted,bend right=-40,latex-] (1.120);

\path (2.300) edge[dotted,bend right=40,-latex] (3.240);
\path (2.60) edge[dotted,bend right=-40,latex-] (3.120);
\end{tikzpicture}
}

\begin{document}

\setlength{\abovedisplayskip}{5pt}
\setlength{\belowdisplayskip}{5pt}

\title{From stochastic spin chains to quantum Kardar-Parisi-Zhang dynamics}

\author{Tony Jin}
\email{zizhuo.jin@unige.ch}
\affiliation{DQMP, University of Geneva, 24 Quai Ernest-Ansermet, CH-1211 Geneva, Switzerland}
\affiliation{Laboratoire de Physique de l'\'Ecole Normale Sup\'erieure, CNRS, ENS $\&$ PSL University, Sorbonne Universit\'e, Universit\'e de Paris, 75005 Paris, France}
\author{Alexandre Krajenbrink}
\email{alexandre.krajenbrink@sissa.it}
\affiliation{Laboratoire de Physique de l'\'Ecole Normale Sup\'erieure, CNRS, ENS $\&$ PSL University, Sorbonne Universit\'e, Universit\'e de Paris, 75005 Paris, France}
\affiliation{SISSA and INFN, via Bonomea 265, 34136 Trieste, Italy}
\affiliation{Cambridge Quantum Computing Ltd, 9a Bridge Street, CB2 1UB Cambridge, United Kingdom}
\author{  Denis Bernard }
\email{denis.bernard@ens.fr}
\affiliation{Laboratoire de Physique de l'\'Ecole Normale Sup\'erieure,  CNRS, ENS $\&$ PSL University, Sorbonne Universit\'e, Universit\'e de Paris, 75005 Paris, France}

\date{\today}

\begin{abstract}
We introduce the asymmetric extension of the Quantum Symmetric Simple Exclusion Process which is a stochastic model of fermions on a lattice hopping with random amplitudes. In this setting, we analytically show that the time-integrated current of fermions defines a height field which exhibits a quantum non-linear stochastic Kardar-Parisi-Zhang dynamics. Similarly to classical simple exclusion processes, we further introduce the discrete Cole-Hopf (or Gärtner) transform of the height field which satisfies a quantum version of the Stochastic Heat Equation. Finally, we investigate the limit of the height field theory in the continuum under the celebrated Kardar-Parisi-Zhang scaling and the regime of almost-commuting quantum noise.
\end{abstract}

\maketitle

Random unitary dynamics arise in quantum mechanics as an efficient way of describing the evolution of systems interacting with environments or external fields. The original idea was introduced by Caldeira and Leggett to study the effective dynamics of collections of spins interacting with bosonic baths \cite{CALDLEGGpathintegraltoQBM}. The properties of such systems are expected to notably differ from their isolated counterparts due to fluctuation and dissipation arising from the interactions with unknown degrees of freedom. Random unitary dynamics are also useful in theoretical studies of \textit{typical and universal} behaviors of quantum chaotic systems. As a consequence, their studies has been recently revitalized, notably in the context of random unitary circuits \cite{Qcircuit01,Qcircuit02,Qrandom-circuit1,Qrandom-circuit2,Qrandom-circuit3,Qcircuit03,Qcircuit04,Qcircuit06}, as well as in conventional many body systems \cite{,Lamacraft,Eisert17,BBJsto,Garrahan17,Knap18,Lama18,CaoTilloyLuca}.
By adding stochasticity, these systems ought to lose their fine properties pertaining to particularities, such as conservation laws, thus allowing the emergence of generic properties. These include the production of entanglement \cite{Qcircuit01,Qrandom-circuit1,EntangHydro01,EntangHydro03,zhou2019emergent,HemeryPollmanLuitz2019,Gullans18,EntangHydro04,EntangHydro02,EntangChaos}, the scrambling of information \cite{Qcircuit00,Qcircuit02,Qrandom-circuit3,Qchaos05}, or the spreading of operators \cite{Qrandom-circuit2,Qcircuit03,Qcircuit04} in systems converging to thermal or out-of-equilibrium steady states. 
In particular, in some quantum stochastic models \cite{Qrandom-circuit1,zhou2019emergent,Knap18,Lama18}, it has been argued that the growth and fluctuations of the entanglement entropies are governed by the Kardar-Parisi-Zhang (KPZ) equation \cite{KPZ,halpin_zhang,krugg,HairerSolvingKPZ,reviewCorwin,HH-TakeuchiReview,TAKEUCHI201877}. Large deviation fluctuations for the growth of entanglement in stochastic conformal field theories have also been shown to belong to the KPZ class \cite{DenisPierre2019}. Some scaling features of the KPZ equation have also recently been found in super-diffusive non-stochastic spin chain models \cite{krajnik2019kardar,das2019kardar,ljubotina2019kardar,dupont2019universal} for the long-time decay of the spin-spin correlation functions. The extent by which the KPZ-like behaviors is universal in quantum many-body systems is still an open question.

In a series of papers \cite{BBJsto,4BBJ,5BJ}, two of the authors and Michel Bauer introduced a stochastic model describing stochastic hoppings of fermions on a lattice which can be seen as a continuous-time version of the random unitary circuit models. In particular, the average dynamics of such quantum model reduces to the classical Symmetric Simple Exclusion Process (SSEP) \cite{5BJ}, a well known model of out-of-equilibrium classical statistical physics. One remarkable results of \cite{4BBJ,5BJ} consisted in showing that, beyond the decoherence effects at work in the mean dynamics, quantum coherences and hence entanglement patterns have a rich structure in such models.
In this Letter, we extend these results to asymmetric processes by introducing an asymmetric version of the quantum SSEP model and deciphering its connection with both the classical Asymmetric Simple Exclusion Process (ASEP) and the classical or quantum KPZ equation. This hence provides one theoretical example of direct appearance of the KPZ physics in quantum spin chains.

The connection between the quantum ASEP model we shall define below, see Eqs.~(\ref{eq:infinitesimal},\ref{eq:dynamique}), and the KPZ dynamics will be declined along different angles. First, we will show that the average dynamics of the model is described by the classical ASEP. Then, we will show how the time-integrated current of fermions follows a quantum stochastic dynamics, akin to the classical KPZ one. This last connection is realized in three different manners: by mapping the spin chain dynamics to a quantum discrete version of the KPZ equation, Eqs.~(\ref{eq:heightfield},\ref{eq:dBdBnoise}), or to a quantum analog of the stochastic discrete heat equation (SHE), Eqs.~(\ref{eq:SHE22},\ref{eq:ZdB}), or to a quantum version of the stochastic Burgers equation, Eq.~\eqref{eq:discreteBurgers}. In all these instances, the noise keeps its quantum character and depends on off-diagonal quantum coherences. Since the classical ASEP and its scaling limit are at the root of the macroscopic fluctuation theory (MFT) \cite{MFT-et-al}, which is an effective theory coding for large deviation fluctuations in diffusive out-of-equilibrium classical systems, the quantum ASEP model (\ref{eq:infinitesimal},\ref{eq:dynamique}) and its mapping to the quantum KPZ dynamics open the route towards the extension of the macroscopic fluctuation theory to quantum coherence and entanglement fluctuations in quantum many body systems.

\begin{figure}
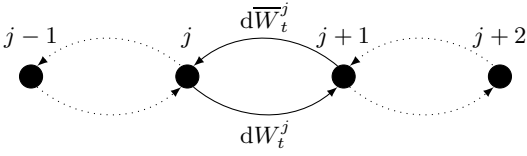

\latticeQuantum
\caption{Representation of the lattice and the hopping amplitudes. A fermionic particle can jump to the right (resp. to the left) with probability per unit time $\rmd W_t$ (resp. $\rmd \overline{W}_t$). These amplitudes correspond to an effective interaction with a bosonic bath. \vspace*{-0.5cm}}
\label{fig:lattice}
\end{figure}

\paragraph{Model and formalism.}
We consider an asymmetric extension of the quantum SSEP model introduced
in \cite{4BBJ}. This model describes fermions on an infinite lattice
undergoing hopping between nearest-neighbour with random amplitudes.
In the Heisenberg picture, the infinitesimal stochastic Hamiltonian
generating the flow on observables $\hat{O}_{t}\mapsto \hat{O}_{t+dt}=e^{i\rmd H_{t}}\hat{O}_{t}e^{-i\rmd H_{t}}$
is given by 
\begin{equation}
\label{eq:infinitesimal}
\rmd H_{t}=\sum_{j=-\infty}^{+\infty}\left[c_{j+1}^{\dagger}c_{j}\rmd W_{t}^{j}+c_{j}^{\dagger}c_{j+1}\rmd \overline{W}_{t}^{j}\right]
\end{equation}
 where $(c_{j}^{\dagger},c_{j})$ are fermionic creation and annihilation
operators acting at site $j$ with usual anti-commutation relation $\lbrace c_j^\dagger, c_i \rbrace=\delta_{ij}$ and $( \rmd W_{t}^{j},\rmd \overline{W}_{t}^{j} )$ are \textit{quantum noises}, attached to the edges, see Fig.~\ref{fig:lattice} for a representation of the process. 
Quantum noises are operators living on the Fock
space ${\cal H}_{{\rm noise}}$ of the reservoir and fulfill the canonical
equal-time commutation relation $[\rmd W_{t}^{j},\rmd \overline{W}_{t}^{k}]=\delta_{jk}\rmd t$.
Physically, they represent operators in the interacting picture creating and annihilating excitations
in a bosonic reservoir. We define the stochastic average as the trace over the degrees of freedom of the bath and denote it $\mathbb{E}[\hspace{1em}]$. Within stochastic averages, the noise satisfies the so-called quantum Itô rules $\rmd \overline{W}_{t}^{j}\rmd W_{t}^{k}=\delta_{jk}\alpha \rmd t$ and $\rmd W_{t}^{k}\rmd \overline{W}_{t}^{j}=\delta_{jk}(1+\alpha)\rmd t$ where $\alpha$ is the average number of bosonic excitations in the bath, and $\mathbb{E}[\rmd W_t^j ]=\mathbb{E}[\rmd \overline{W}_t^j]=0$. Our model further assumes that the noise is Markovian in the sense that there is no memory effect due to the large number of degrees of freedom of the bosonic bath: this implies $\rmd \overline{W}_{t}^{j}\rmd W_{t'}^{k}=\rmd W_{t}^{k}\rmd \overline{W}_{t'}^{j}=0$ for $t\neq t'$. 
We refer the reader to the existing literature \cite{Qnoise1,Qnoise2,BBTOpenQBM,Attal2006} for more details on quantum noise.

The equation of evolution of a generic operator $\hat{O}$ follows from the expansion of the flow \eqref{eq:infinitesimal} of $\hat{O}_{t}$ up to the second order in $\rmd H_t$, \cite{footnoteLeftRight},
\begin{align}
\rmd\hat{O}_{t} & =\left[\mathcal{L}_{\mathrm{TASEP}}^{*}(\hat{O}_{t})+\alpha \mathcal{L}_{\mathrm{SSEP}}^{*}(\hat{O}_{t})\right]{\rm d}t\label{eq:dynamique}\\
 & \hspace{1em}+i\sum_{j=-\infty}^{+\infty}[c_{j+1}^{\dagger}c_{j}\mathrm{d}W_{t}^{j}+c_{j}^{\dagger}c_{j+1}\mathrm{d}\overline{W}_{t}^{j},\hat{O}_{t}]\, . \nonumber 
\end{align}
The super-operators are given here by $\mathcal{L}_{\mathrm{TASEP}}^{*}(\star)=\sum_{j=-\infty}^{+\infty}[c_{j+1}^{\dagger}c_{j}\star c_{j}^{\dagger}c_{j+1}-\frac{1}{2}\lbrace(1-\hat{n}_{j})\hat{n}_{j+1},\star\rbrace]$
 and $\mathcal{L}_{\mathrm{SSEP}}^{*}(\star)=\sum_{j=-\infty}^{+\infty}[c_{j}^{\dagger}c_{j+1}\star c_{j+1}^{\dagger}c_{j}+c_{j+1}^{\dagger}c_{j}\star c_{j}^{\dagger}c_{j+1}-\frac{1}{2}\lbrace\hat{n}_{j}(1-\hat{n}_{j+1})+(1-\hat{n}_{j})\hat{n}_{j+1},\star\rbrace]$. 
The notation $\{,\}$ stands for the anti-commutator
and $\hat{n}_{j}$ is the local density $c_{j}^{\dagger}c_{j}$. The superscript
$^{*}$ denotes the fact that we are working with the dual of a Lindbladian
on the operator space and we will explain the meaning of the subscript in the following. As an example, the dual Lindlabians evaluated on the number operator $\hat{n}_k$ yield 
\begin{equation}
\begin{split}
   \mathcal{L}^*_{\mathrm{TASEP}}(\hat{n}_k)   &= \hat{n}_{k+1}(1-\hat{n}_{k}) -\hat{n}_{k}(1-\hat{n}_{k-1}) \, ,\\
    \mathcal{L}^*_{\mathrm{SSEP}}(\hat{n}_k)    &= (\hat{n}_{k+1}-\hat{n}_k)-(\hat{n}_k-\hat{n}_{k-1})\, .
    \end{split}
    \label{eq:LindlabialNumber}
\end{equation}
The form of these equations echoes the fermion number conservation, which leads to the local continuity equation, $\rmd \hat{n}_k= (j_k\rmd t+ \rmd B_t^k)- (j_{k-1} \rmd t+ \rmd B_t^{k-1})$, with noise
\begin{equation}
\rmd B^k_t=i[c_{k+1}^{\dagger}c_{k}\mathrm{d}W_{t}^{k}-c_{k}^{\dagger}c_{k+1}\mathrm{d}\overline{W}_{t}^{k}]\, ,
\label{eq:noise}
\end{equation}
and current $j_k = \big(\hat{n}_{k+1}(1-\hat{n}_{k})  + \alpha (\hat{n}_{k+1}-\hat{n}_k)\big)$. Since these Lindblad operators act diagonally on the number operators, they encode for a Markovian master equations. 

\paragraph{Correspondance to the ASEP.}

The mean dynamics of this model can be mapped to classical exclusion processes. Let us define pointer-states as elements of the form $\mathbb{P}_{[\boldsymbol{\epsilon}]}=\prod_{j}^{\otimes}\mathbb{P}_{j}^{\epsilon_{i}}$
with $\boldsymbol{\epsilon}\equiv\{\epsilon_{j}\}$, $\epsilon_{j}\in\{\pm\}$ and
$\mathbb{P}_{j}^{+}=c_{j}^{\dagger}c_{j}$, $\mathbb{P}^{-}=(\mathbb{I}-c_{j}^{\dagger}c_{j})$.
These operators are sent onto each others by the generator of the mean evolution: $\mathbb{E}[{\cal L}(\mathbb{P}_{[\boldsymbol{\epsilon}]})]=\sum_{\boldsymbol{\epsilon}'}Q[\boldsymbol{\epsilon'}]\mathbb{P}_{[\boldsymbol{\epsilon}']}$. One then associates to each pointer-state a classical state such that $\mathbb{P}_{j}^{+}$ (resp. $\mathbb{P}_{j}^{-}$) corresponds to an occupied state (resp. empty state) at site
$j$ which we denote $|\full|$ (resp. $|\empt|$). Starting from a diagonal density matrix $\rho_{0}=\sum_{\boldsymbol{\epsilon}}Q_{0}[\boldsymbol{\epsilon}]\mathbb{P}_{[\boldsymbol{\epsilon}]}$,
the dynamics leaves the density matrix $\rho_t$ diagonal on average and the weights $Q_{t}[\boldsymbol{\epsilon}]$ at time $t$
can be associated to the classical probability amplitude for the system to be in the configuration $\boldsymbol{\epsilon}$. The master equation
satisfied by the weights $Q_{t}$ can be found upon evaluation of the
action of the Lindbladian on a pair of adjacent sites: 
\begin{align*}
{\cal L}_{{\rm SSEP}}(|\fullempt|) & =-|\fullempt|+|\emptfull| ,\\
{\cal L}_{{\rm SSEP}}(|\emptfull|) & =-|\emptfull|+|\fullempt| ,\\
{\cal L}_{{\rm TASEP}}(|\fullempt|) & =0 ,\\
{\cal L}_{{\rm TASEP}}(|\emptfull|) & =-|\emptfull|+|\fullempt|.
\end{align*}
Moreover, both operators return zero upon acting on the states $|\emptempt|$ and $|\fullfull|$ since either there is no particle or the exclusion freezes the dynamics.
The transition rates are thus equal to the ones for the corresponding
classical processes: ${\cal L}_{{\rm SSEP}}$ corresponds to the transition
rates of the symmetric simple exclusion process while ${\cal L}_{{\rm TASEP}}$
corresponds to the totally asymmetric one. We emphasize that these statements are only true if we restrict ourselves to the average evolution and consider pointer-states initial conditions. This quantum model embraces much more general situations where quantum coherences play a key role both from the initial condition and the dynamics \cite{4BBJ,5BJ}. At large scale, the classical ASEP is known to converge towards the KPZ growth model under a weak asymmetry. In what follows, we show that a similar statement holds for the quantum case.

\begin{figure}
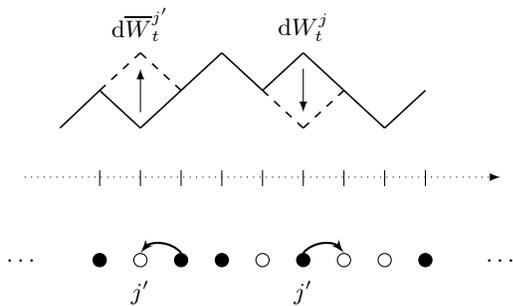

\lattice
\caption{Classical mapping from the particle picture to the height field. \textbf{Bottom.}  The particles can hop on the lattice, the occupied sites are black-filled and the rest is empty. \textbf{Top.} The configuration of particles can be represented by an interface incremented with the presence of a particle at the corresponding site. The particle transport maps to a corner growth process: if a particle hops to the left (resp. to the right) then a corner is raised (resp. lowered). \vspace*{-0.5cm}}
\label{fig:mappingHeight}
\end{figure}

\paragraph{Introduction of the Kardar-Parisi-Zhang quantum height field.} The remaining of this Letter consists in introducing the Kardar-Parisi-Zhang physics in the context of our quantum model. For our purpose, let us recall that it is described by a trio of equations: the KPZ equation, the stochastic heat equation and the stochastic Burgers equation. Our first task consists in unveiling their quantum counterpart in our model. To this aim, 
let $\tilde{n}_{j}\equiv n_{j}-\frac{1}{2}$ be the centered local
density and $\hat{h}_{k}\equiv\sum_{j=-\infty}^{k}\tilde{n}_{j}$ be  the cumulated fermion number, or identically the time-integrated current of fermions \cite{footnoteIntegrated}, to which we will now refer as the \textit{quantum height field}.
In classical exclusion processes, the height is defined similarly as an integrated current and its growth is related to the so-called \textit{corner growth process} consisting in the growth of an interface from its corners, transforming local minima (resp. maxima) onto local maxima (resp. minima) whenever a particle passes through the corresponding site, see Fig.~\ref{fig:mappingHeight}. 
From now on and for convenience, we introduce the forwards gradient of the height as $\nabla_+ \hat{h}_k=\hat{h}_{k+1}-\hat{h}_k$, its backwards gradient as $\nabla_- \hat{h}_k=\hat{h}_{k}-\hat{h}_{k-1}$ and the discrete Laplacian as $\Delta=\nabla_+ \nabla_-$.
From Eq.~\eqref{eq:dynamique} or from the summation of the evolution of the number operators $\hat{n}_j$, we obtain the exact dynamics of the height as 
\begin{align}
\mathrm{d}\hat{h}_{k}  =[(\alpha+\frac{1}{2})\Delta\hat{h}_{k}-\nabla_+ \hat{h}_k \nabla_- \hat{h}_k+\frac{1}{4}]\mathrm{d}t + \rmd B_t^k.
 \label{eq:heightfield}
\end{align}
Note that the non-linearity is actually a direct consequence of the non-commutativity of the noise.
We emphasize that the noise term \eqref{eq:noise} depends on \textit{off-diagonal observables} with respect to the occupation number basis: therefore the fluctuations of the height, contrary to the classical case, is strongly dependent on the existence of quantum coherences. 
Let us comment the evolution \eqref{eq:heightfield} of the height: the deterministic part of the evolution comprises a non-linear term
rendering the equation intrinsically difficult to solve and the correlator of the noise depends on the height as
\begin{equation} \label{eq:dBdBnoise}
 \rmd B_t^k \rmd B_t^{k'} =\delta_{kk'}[ \frac{1}{2}\Delta \hat{h}_k+(\alpha+\frac{1}{2}) (\frac{1}{2}-2\nabla_+ \hat{h}_k\nabla_- \hat{h}_k)]\mathrm{d}t
\end{equation}
Equation \eqref{eq:heightfield} plays the role of a discretized quantum KPZ equation. To further study the model, let us remark that there exists a \textit{miracle}: if one defines the operator $\hat{Z}_{k}\equiv e^{\delta\hat{h}_{k}}$, then for a suitable choice of parameter $\delta$, the dynamics of $\hat{Z}$ will become linear. This transformation is similar to the so-called discrete Cole-Hopf (or Gärtner) change of variable for classical models \cite{Gartner1, Gartner2, Bertini.Giacomin-CMP1997} and for this reason we will refer to $\hat{Z}$ as the quantum Cole-Hopf (or Gärtner) operator.     From an explicit application of Eq.~\eqref{eq:dynamique} or from Eq.~\eqref{eq:heightfield}, the evolution of the Cole-Hopf operator \cite{footnoteLaplacian} is given by
\begin{align}
\rmd \hat{Z}_{k} & =\big[(1+\alpha)(1-\hat{n}_{k})\hat{n}_{k+1}e^{\delta/2}-\alpha\hat{n}_{k}(1-\hat{n}_{k+1})e^{-\delta/2}\big] \nn \\
&\hspace{1em}\times 2\sinh(\frac{\delta}{2})\hat{Z}_{k}\rmd t + 2\sinh(\frac{\delta}{2})\, \rmd B_t^{k,(-)}\hat{Z}_{k}\, ,
\label{eq:SHEdiscrete}
\end{align}
where we further introduced the decorated noise $ \mathrm{d}B_t^{k,(\pm)}=i[ e^{\pm\frac{\delta}{2}} c_{k+1}^\dagger c_k  \mathrm{d}W_t^k- e^{\mp\frac{\delta}{2}} c_k^\dagger c_{k+1} \mathrm{d}\overline{W}_t^k ]$. We can see that Eq.~\eqref{eq:SHEdiscrete} remains non-linear in the number operators for an arbitrary parameter $\delta$. Nonetheless, a wise choice allows to cancel the non-linearity $\delta=\log(\alpha/(1+\alpha))$. Fixing this  value, the evolution of the Cole-Hopf operator is simplified and reads
\begin{align}
\rmd \hat{Z}_{k} & =\left[\sqrt{\alpha(\alpha+1)}\Delta \hat{Z}_{k}-(\sqrt{\alpha+1}-\sqrt{\alpha})^2\hat{Z}_{k}\right]\rmd t \nn \\
&\hspace{1em}-\frac{1}{\sqrt{\alpha(\alpha+1)}} \, \rmd B_t^{k,(-)}\hat{Z}_{k}.
\end{align}
The constant drift can be absorbed by redefining the Cole-Hopf operator as $\tilde{Z}_{k}\equiv e^{\mu t}\hat{Z}_{k}$
with $\mu=(\sqrt{\alpha +1}-\sqrt{\alpha })^2$. This amounts to shift the height field in another frame in translation with the original one. Upon this change, the evolution of $\tilde{Z}_k$ satisfies a discrete quantum stochastic heat equation 
\begin{equation}
\hspace*{-0.35em}\rmd \tilde{Z}_{k}  = \sqrt{\alpha(1+\alpha)}\Delta\tilde{Z}_{k}\rmd t-\frac{1}{\sqrt{\alpha(\alpha+1)}}  \rmd B_t^{k,(-)}\tilde{Z}_k.  
\label{eq:SHE22}
\end{equation}
Furthermore, let us note that from the specific structure of the decorated noise, it is possible to rearrange the noise term for convenience in three different ways
\begin{equation} \label{eq:ZdB}
\tilde{Z}_{k}^{1/2}\mathrm{d}B_t^k \tilde{Z}_{k}^{1/2}=\tilde{Z}_{k}\mathrm{d}B_t^{k,(+)}=\mathrm{d}B_t^{k,(-)}\tilde{Z}_k \,.
\end{equation}
We see that going from Eq.~\eqref{eq:heightfield} to Eq.~\eqref{eq:SHE22} we have exchanged one difficulty with another one: we have traded a non-linear equation and an additive noise with a linear equation with a multiplicative noise. A first remark on the discrete SHE is that the stochastic averaged dynamics of the Cole-Hopf operator can be computed exactly since it describes a diffusion on $\Z$ \cite{ciaurri2017harmonic}
\begin{equation}
\hspace*{-0.69em}    \mathbb{E}[\tilde{Z}_k(t)]= \sum_{\ell\in \mathbb{Z}} I_{k-\ell}(2 t\sqrt{\alpha(\alpha+1)}) e^{-2 t\sqrt{\alpha(1+\alpha)}}\tilde{Z}_\ell(0) 
    \label{eq:averageZ}
    \vspace*{-0.05cm}
\end{equation}
where $I$ is the modified Bessel function of the first kind \cite{abramowitz1948handbook}. We emphasize that Eq.~\eqref{eq:averageZ} is operator-valued since the trace over the fermionic degrees of freedom has not been taken. Finally, the third equation of the trio is obtained quite readily. Indeed, the backwards gradient of the quantum height field, $\tilde{n}_k=\nabla_- \hat{h}_k$ verifies a quantum discrete viscous stochastic Burgers equation 
\begin{equation}
\rmd \tilde{n}_k+\tilde{n}_k(\nabla_++\nabla_-)\tilde{n}_k\rmd t=(\alpha+\frac{1}{2})\Delta \tilde{n}_k \rmd t+\nabla_- \rmd B^k_t \, .
\label{eq:discreteBurgers}
\end{equation}

It is classically more convenient to study the stochastic Burgers equation to obtain stationary measures in the KPZ physics, the reason for that is that one expects the height to have a linear time growth preventing stationarity. The gradient of the height will not show a constant growth and will therefore be the right quantity to investigate stationarity: we expect the same phenomenon to remain valid in the quantum realm.  Now that we have unveiled the trio of equation governing the KPZ physics, let us now turn to the introduction of quantum replica of the Cole-Hopf operator and to the limit of our field theory in the continuum.

\paragraph{Quantum replica.} A possible direction to study the stochastic heat equation is by the means of the replica method pioneered by Kardar in \cite{kardareplica}. We propose in this Letter to extend it to our quantum model. To this aim, we define the equal-time $n$-th quantum replica as \, $u(k_1,\dots,k_n)= \prod_{\ell=1}^n \tilde{Z}_{k_\ell}$, 
which evolution can be derived from the Itô rules and the discrete SHE \eqref{eq:SHE22} as 
\begin{equation}
\partial_t \mathbb{E}[u]\hspace*{-0.5em}   =  \mathbb{E} \big[\sqrt{\alpha(\alpha+1)} \sum_{\ell=1}^n \Delta_\ell  \; u +  \sum_{\ell<m} V_{k_\ell,k_m} u  \big] 
    \label{eq:LiebLiniger}
\end{equation}
The potential $V$ is given by the correlator of the noise
\begin{equation}
\begin{split}
&V_{k_\ell,k_m} = \frac{1}{\alpha(\alpha+1)} \frac{\mathrm{d}B_t^{k_\ell,(+)} \mathrm{d}B_t^{k_m,(-)}}{\mathrm{d}t}\\
    &= \delta_{k_mk_\ell} [ \frac{(1-\hat{n}_{k_m})\hat{n}_{k_m+1}}{1+\alpha}+\frac{ \hat{n}_{k_m}(1-\hat{n}_{k_m+1})}{\alpha}] \, .
    \end{split}
\end{equation}
Equation~\eqref{eq:LiebLiniger} describes an imaginary-time Schrödinger equation on a lattice acting on a space of operators and shares some strong similarities with the $\delta$-Bose gas or the Lieb-Liniger model \cite{ll,ll2}. Indeed, the potential $V_{k_\ell,k_m}$ is an attractive contact interaction and since all operators $Z_{k_\ell}$ commute, $u(k_1,\dots,k_n)$ is invariant by permutation of the $k$'s and therefore each replica can be interpreted as a bosonic particle. The Lieb-Liniger model is solvable using the Bethe ansatz \cite{gaudin2014bethe}, it would be of interest to develop an extension of this ansatz for our Hamiltonian \eqref{eq:LiebLiniger} acting on the space of operators. We leave the development of such ansatz for a future work.

\paragraph{Limit to the continuum -- (1:2:3) scaling and almost-commuting quantum noise.} So far, our quantum model was considered on an infinite lattice. To investigate its large scale properties, we propose to probe its limit in the continuum in a similar fashion as for the classical exclusion processes \cite{Bertini.Giacomin-CMP1997}. We will show that in a regime that we call the \textit{almost-commuting quantum noise regime} and in the celebrated (1:2:3) KPZ scaling limit, the continuous version of Eq.~\eqref{eq:heightfield}
is a quantum version of the continuous Kardar-Parisi-Zhang equation. The (1:2:3) KPZ scaling \cite{footnoteScaling} refers to the following (height : space : time) scaling 
\begin{equation}
\lbrace
\hat{h}\equiv\varepsilon^{-1} \tilde{h}, k \equiv \lfloor\varepsilon^{-2}x \rfloor ,t\equiv\varepsilon^{-3}\tilde{t} \rbrace \, .
\label{eq:KPZscaling}
\end{equation}  
In addition, the almost-commuting quantum noise regime is defined as the limit of an almost infinite number of excitations in the bosonic bath
\begin{equation}
\alpha\equiv\varepsilon^{-1} \alpha_0\, .
\end{equation}
Both regimes are then considered in the limit $\varepsilon\to0$. At leading order in $\varepsilon$, the (1:2:3) scaling leads to the following minimal replacements $ \Delta h \to \varepsilon^3\partial_{xx} \tilde{h}$, $
        \nabla h  \to \varepsilon\partial_{x} \tilde{h}$ and $
        \delta_{kk'}\to \varepsilon^2\delta(x-x')$. The equation of evolution of the quantum height \eqref{eq:heightfield} is then transformed onto
\begin{equation}
\mathrm{d}\tilde{h}=\big[\alpha_{0}\partial_{xx}\tilde{h}-(\partial_{x}\tilde{h})^{2}+\frac{1}{4\varepsilon^{2}}\big]\mathrm{d}\tilde{t}+\mathrm{d}B_{\tilde{t}}^{x}\, ,
\label{eq:KPZ}
\end{equation}
where the noise in continuum is defined as $ \rmd B_{\tilde{t}}^{x}\equiv\lim_{\varepsilon\to0} \varepsilon  \,   \rmd B_t^k$.
The correlator of the noise in the continuum is extremely simple and is purely classical at leading order in $\varepsilon$
\begin{equation}\label{eq:Itoclassic}
\begin{split}
& \mathrm{d}B_{\tilde{t}}^{x}\mathrm{d}B_{\tilde{t}}^{x'}
 \underset{\varepsilon\to0}{\longrightarrow}\delta(x-x')\frac{\alpha_{0}}{2}\mathrm{d}\tilde{t}\, . 
  \end{split}
\end{equation}
Let us comment the continuous equation \eqref{eq:KPZ} we just obtained. It is extremely similar to the classical KPZ equation although being operator-valued. In addition, similarly to what happens classically \cite{Bertini.Giacomin-CMP1997}, since we expect the quantum interface described by the height $\tilde{h}$ to be \textit{rough}, the $\varepsilon^{-2}$-divergence in Eq.~\eqref{eq:KPZ} is not surprising: one needs to renormalize the field by the addition of a correction term $t/4\varepsilon^2$ to obtain a well-defined theory. It is remarkable that the quantum character of the noise disappears in the continuum limit, see Eq.~\eqref{eq:Itoclassic}. Nonetheless, let us note that the non-linearity remains relevant: since it originated from the non-commutativity of the quantum noise, this is a striking evidence that the model conserves some of its quantum properties on large scales. The discrete SHE \eqref{eq:SHE22} and the discrete Burgers equation \eqref{eq:discreteBurgers} converge in the same fashion to the continuous SHE and stochastic Burgers equation.

There exist a number of analytic results concerning the classical KPZ equation in 1+1 dimensions. Explicit solutions have been found for a variety of initial conditions in full-space and also on the half-line \cite{SS10,CLR10,dotsenko,ACQ11,CLDflat,SasamotoStationary,BCFV,borodin2016directed,barraquand2017stochastic,deNardisPLDTT,KrajenbrinkReplica,KrajenbrinkHalfSpace2,krajenbrink2018large,gueudre2012directed}, it would be interesting to know whether these admit an analog in the quantum case. Furthermore, the classical stochastic Burgers equation admits a Gaussian stationary measure \cite{SasamotoStationary,prahofer2004exact122} with the correlator given on the r.h.s of \eqref{eq:correlator}. This hints at the existence of a stationary measure in the quantum case with a correlator for the quantum height gradient $\partial_x \tilde{h}$ (or equivalently the fermion density-density correlator) given by the classical correlator 
\begin{equation}
\mathbb{E}\left[\langle \partial_x \tilde{h}(0,0)\partial_x\tilde{h}(x,\tilde{t})\rangle\right]  \sim \frac{1}{ \tilde{t}^{2/3}} f_{\rm KPZ}\bigg(\frac{x}{2 \tilde{t}^{2/3}}\bigg)\, , \vspace*{-0.1cm}
\label{eq:correlator}
\end{equation}
where $f_{\rm KPZ}(x)$ was calculated in \cite{prahofer2004exact122,SasamotoStationary}. It is positive, symmetric, normalized to unity and decays as $\sim \exp(-c \vert x\vert^3)$. An important question at this stage would be to investigate the stationary measure of the quantum Burgers equation. We leave this for a future work.

\paragraph{Discussion and outlook.} 
In this Letter, we have introduced a quantum Asymmetric Simple Exclusion Process and identified a quantum growing interface $\hat{h}(t)$ sharing strong similarities with a classical counterpart: the Kardar-Parisi-Zhang height. The KPZ physics is classically described by a trio of equations: the KPZ equation itself, the stochastic Burgers equation and the stochastic heat equation and we have established here the existence of their quantum counterpart both in the discrete and the continuous settings. This model provides one analytic example of the presence of the KPZ physics in a quantum model and we hope it will stimulate the community and help bridge the gap between classical and quantum out-of-equilibrium physics.
	
This model raises nonetheless a number of questions. From a technical point of view it would be interesting to pursue the study of the moments of the Cole-Hopf operator and also investigate the full counting statistics  and the entanglement entropy in this model. Besides, although the infinitesimal Hamiltonian \eqref{eq:infinitesimal} is quadratic in the fermionic operators, common techniques such as Wick's theorem fail here due to the quantum nature of the hopping amplitudes, it would be of great importance to develop new tools to determine whether this model is solvable. The results \cite{5BJ} about the partner quantum SSEP model seems to point toward an underlying integrable structure. Understanding better the role of the off-diagonal quantum coherences in the model behavior would also be important. From a conceptual point of view, it would be interesting to obtain the stationary measure of the theory. 

As an outlook, let us mention that this model further questions the existence of a quantum KPZ-like universality class. There are in the classical Kardar-Parisi-Zhang physics two major results concerning universality. The first one is the strong universality which states that for growth models with a fixed asymmetry, there exists a fixed-point at large-time called the KPZ fixed-point \cite{KPZFixedPoint,quastel2017totally,matetski2016kpz} which describes among other quantities the one-point fluctuations of the height with the celebrated $1/3$ exponent $h(t)\sim v_\infty t + t^{1/3} \chi$, where $v_\infty$ is the average velocity of the interface and $\chi$ is a random variable depending on some large scale features of the initial condition belonging to a universal family. The second result is the weak universality \cite{HairerQuastel2015} which asserts that the KPZ equation is the universal scaling limit of weakly-asymmetric growth models under the (1:2:3) scaling \eqref{eq:KPZscaling}. Extending these results to the quantum realm is one challenge that could shed some new light on quantum out-of-equilibrium problems. 

Finally, solving the SSEP and ASEP models have been instrumental in the formulation of the macroscopic fluctuation theory (MFT) \cite{MFT-et-al}. We may expect that solving exactly their quantum versions introduced in the Letter will play an analogue role in the formulation of a quantum extension of the MFT, aiming at describing quantum coherence and entanglement fluctuations in out-of-equilibrium quantum many body systems and completing the recently emerging membrane picture \cite{EntangHydro03,zhou2019emergent,EntangHydro02,EntangHydro04} for entanglement production in many-body systems.

\acknowledgments 
\paragraph{Acknowledgments.}  We thank M. Bauer, P. Le Doussal and T. Gautié for very helpful discussions. AK acknowledges support from the ANR grant ANR-17-CE30-0027-01 RaMaTraF and from ERC under Consolidator grant number 771536 (NEMO). TJ acknowledges support from the Swiss National Science Foundation, Division II. TJ and DB acknowledge support from the ANR grant ANR-14-CE25-0003.

\newpage{\pagestyle{empty}\cleardoublepage}


\begin{thebibliography}{100}
\bibitem{footnoteLeftRight} We have chosen here a decomposition in terms of $\alpha$ but we could identically describe the evolution by the sum of two super-operators describing the hopping of the particles to the left and to the right on the lattice $\alpha \mathcal{L}^*_{\rm right}+(1+\alpha)\mathcal{L}^*_{\rm left}$ with $\mathcal{L}_{\mathrm{SSEP}}^{*}=\mathcal{L}_{\mathrm{left}}^{*}+\mathcal{L}_{\mathrm{right}}^{*}$ and $\mathcal{L}_{\mathrm{TASEP}}^{*}=\mathcal{L}_{\mathrm{left}}^{*}$.
\bibitem{footnoteIntegrated} Since the number operators verify a continuity equation $\rmd \hat{n}_k= (j_k\rmd t+ \rmd B_t^k)- (j_{k-1} \rmd t+ \rmd B_t^{k-1})$, one can integrate it with respect to the space and time variables to obtain that $\hat{h}_k=\int_0^t [j_k\rmd t+ \rmd B_t^k]$ and therefore the height field also has the interpretation of a time-integrated current.
\bibitem{footnoteLaplacian} Note that since we work with operators, usual chain rules do not apply. For instance, since the number operators verify $\hat{n}=\hat{n}^2$, we have the following relation between the Laplacians of the height and of the Cole-Hopf operator $
     \Delta \hat{Z}_k=\hat{Z}_k(( e^{\frac{\delta}{4}}- e^{-\frac{\delta}{4}})^2+\Delta \hat{h}_k(e^{\frac{\delta}{2}}-e^{-\frac{\delta}{2}}))$.
\end{thebibliography}
\end{document}